\documentclass{emulateapj}

\slugcomment{The Astrophysical Journal, 679: ? - ?, 2008 June 1
(scheduled)}

\shorttitle{NGC1600}
\shortauthors{Smith, R.M. et al.}

\begin{document}

\title{NGC1600 - Cluster or Field Elliptical?}

\author{Rodney M. Smith\altaffilmark{1},
Vicent J. Mart\'{\i}nez\altaffilmark{2,3}, 
Alberto Fern\'andez-Soto\altaffilmark{3}, 
Fernando J. Ballesteros\altaffilmark{2}, and \\
Amelia Ortiz-Gil\altaffilmark{2}}

\email{Vicent.Martinez@uv.es}

\altaffiltext{1}{Dept. of Physics and Astronomy, Cardiff University,
Cardiff, CF24 3AA, U.K.}
\altaffiltext{2}{Observatori Astron\`omic, Universitat de Val\`encia,
Apartat de Correus 22085, E-46071 Val\`encia, Spain}
\altaffiltext{3}{Departament d'As\-tro\-no\-mia i As\-tro\-f\'{\i}\-si\-ca, Universitat
de Val\`encia, Apartat de Correus 22085, E-46071 Val\`encia, Spain}

\begin{abstract}
A study of the galaxy distribution in the field of the elliptical
galaxy NGC1600 has been undertaken. Although this galaxy is often
classified as a member of a loose group, all the neighbouring galaxies
are much fainter and could be taken as satellites of NGC1600. The
number density profile of galaxies in the field of this galaxy shows a
decline with radius, with evidence of a background at approximately
1.3 Mpc. The density and number density profile are consistent with
that found for other isolated early-type galaxies. NGC1600 appears as
an extended source in X-rays, and the center of the X-ray emission
seems not to coincide with the center of the galaxy. The velocity
distribution of neighbouring galaxies has been measured from optical
spectroscopic observations and shows that the mean radial velocity is
approximately $85 \rm{\ km\ s}^{-1}$ less than that of NGC1600,
indicating that the centre of mass could lie outside the galaxy.  The
velocity dispersion of the `group' is estimated at $429 \pm 57 \rm{\
km\ s}^{-1}$. The inferred mass of the system is therefore of the
order of $10^{14} M_\odot$, a value that corresponds to a large
group. NGC1600 therefore shares some similarities, but is not
identical to, the `fossil clusters' detected in X-ray surveys.
Implications of this result for studies of isolated early-type
galaxies are briefly discussed.

\end{abstract}

\keywords{galaxies: elliptical and lenticular }

\section{Introduction}

The importance of isolated early-type galaxies in understanding galaxy and
cluster evolution has led to a surge in the available number of catalogues
of such objects (e.g Colbert et al. 2001, Smith et al. 2004, Reda et
al. 2004, Denicolo et al. 2005, Buote 2005), with few galaxies in common
between them. The variation arises from the different selection criteria
used in the catalogue construction.  A common requirement is that there is
no galaxy with a similar redshift within a certain distance, typically of
the order of 1 Mpc. However, the incompleteness of current galaxy
catalogues, and in particular the lack of complete redshift information,
usually requires a visual inspection of the field as final confirmation
that the galaxies are isolated. The catalogue of Smith et al. (2004,
hereafter SMG) only used redshift information for the central galaxy and
discounted any galaxies which had a bright galaxy nearby in projection. It
thus contains the strictest isolation criteria and all galaxies in the
sample are the most dominant member in the field by at least 2.2 magnitudes
within 500 kpc and 0.7 magnitudes within 1 Mpc. Again, visual inspection
confirmed that these galaxies were indeed isolated. The SMG catalogue will
miss several galaxies in the other samples as no redshift information is
applied. However, interpretation of the local faint galaxy environment is
simplified by the lack of other bright galaxies nearby which may have their
own dwarf population.

\object{NGC1600}, an 11.9 $B$-magnitude ($K$=8.1\footnote{All near-infrared
magnitudes in this paper have been obtained from the 2MASS
catalogues.}) E3-E4 gal\-axy at a distance of approximately 64 Mpc
(NED\footnote{The NASA/IPAC Extragalactic Database (NED) is operated
by the Jet Propulsion Laboratory, California Institute of Technology,
under contract with the National Aeronautics and Space
Administration.} and using $H_0 = 73 \rm{\ km\ s}^{-1}\rm{Mpc}^{-1}$)
was included in the sample of SMG due to the large magnitude
difference between it and other nearby galaxies, even though it is
often quoted as being at the centre of a loose group containing a
number of NGC-catalogued galaxies together with several other fainter
galaxies. Denicolo et al. (2005) classify it as a field elliptical
rather than isolated.  The brightest galaxy within a projected
distance of 1 Mpc is NGC1594, a 13.9 $B$-magnitude ($K$=10.2) galaxy
at least 800 kpc away. The two closest NGC galaxies, NGC1601 and
NGC1603 at separations of 1.6 and 2.6 arcminutes (30 kpc and 48 kpc)
respectively, are both almost three magnitudes fainter than
\object{NGC1600}. Thus, although there are several neighbouring galaxies in the
field of \object{NGC1600}, it is by far the dominant galaxy in the group. There
is a poor group of galaxies (Zwicky cluster 0430.8-0424B, Zwicky et
al. 1961-1968) at a distance of 53 arcmin to the North of \object{NGC1600},
with a mean velocity of 5030 $\rm{km\ s}^{-1}$ (Baiesi-Pillastrini et
al. 1984), a velocity similar to that of \object{NGC1600} (4715 $\rm{km\
s}^{-1}$).  The brightest galaxy in this group still satisfies the
isolation criteria of SMG, i.e.  it is more than 2.2 magnitudes
fainter than \object{NGC1600}.

The galaxy has been the subject of several previous studies. In a
photometric study, Matthias and Gerhard (1999) found the galaxy to
have boxy isophotes, whilst kinematic studies (Bender et al. 1994,
Faber et al. 1989) found that there was little rotation in the stellar
component, with a maximum of 30 $\rm{km\ s}^{-1}$, whilst the central
velocity dispersion is typical of larger elliptical galaxies (321
$\rm{km\ s}^{-1}$). There is evidence of past and possible ongoing
star formation, with the presence of H$\alpha$ regions (Trinchieri and
di Serego Alighieri 1991) and dust (Ferrari et al. 1999).  Thus it is
highly likely that \object{NGC1600} is a merger remnant (Matthias and Gerhard
1999) although, with broadband colours typical of elliptical galaxies
(NED), then this merger is likely to have occurred a long time ago.
Estimates of the age of this galaxy hence range from 4.6 Gyr up to 8.8
Gyr (Trager et al. 2000, Terlevich and Forbes 2002).These results are
in agreement with the study of a sample of isolated early-type
galaxies by Reda et al. (2004), who found that several galaxies had
boxy isophotes, evidence of dust and past merger activity but had
remained relatively dormant for several Gigayears.

\object{NGC1600} is also a weak X-ray source (Sivakoff et al. 2004), showing
extended emission out to at least 100 arcsec (corresponding to a
physical radius of 30 kpc , with a possible central component
associated with \object{NGC1600} and an outer emission region associated with
the group. The extended outer emission is centred to the North-East of
the central galaxy, suggesting that the centre of the potential is
slightly offset from \object{NGC1600}. The presence of extended material around
the galaxy is also suggested from the tailed X-ray structure around
the nearby galaxy NGC1603, indicating an effect of ram-pressure
stripping. The large magnitude difference between \object{NGC1600} and the
other galaxies in its neighbourhood, together with the extended X-ray
emission, would suggest that \object{NGC1600} may be a fossil group. 

Whereas the radius of about 200 arcsec, or 60 kpc, total soft X-ray
emission of $2 \times 10^{41} \rm{erg\ s}^{-1}$ and absolute magnitude
of $M_R = -22.5$ indicate that this galaxy does not satisfy the
criteria to be a fossil group, as stated by Jones et al. (2003), it is
worthwhile remarking that in recent work by Santos et al (2007) some
cases of fossil groups are identified which resemble some of the
\object{NGC1600} group properties. Moreover, detailed studies by Mendes de
Oliveira et al. (2006) and Cypriano et al (2006) have measured
properties of fossil groups that make them more similar to clusters
than was initially expected. To distinguish between the various
possible evolutionary scenarios (rich group, fossil group or isolated
galaxy) a more detailed investigation of the potential around \object{NGC1600}
is thus necessary.

In a similar technique to that used by Zaritsky et al. (1993,1997), it
is possible to use the dynamics of the fainter galaxies to derive an
estimate of the size and mass of the central potential.  This should
enable us to distinguish between the isolated or cluster hypotheses
for the description of \object{NGC1600}. From the dynamics of galaxies in
groups, estimates of their mass are typically of the order of
$10^{13}M_\odot$ (e.g. Parker et al., 2005) whilst spiral galaxies
have masses an order of magnitude smaller, even allowing for any dark
matter component at large radii (e.g. Zaritsky et al. 1997). The radii
of groups is also an order of magnitude greater than that of the
individual galaxies themselves (e.g. Karachentsev 2005). The mass and
extent of elliptical galaxies is at present very uncertain, with X-ray
evidence that some are surrounded by dark matter haloes (e.g. Fukazawa
et al. 2006) whilst optical evidence may suggest not, with masses
similar to that of spiral galaxies (e.g.  Romanowsky et al. 2003).
However, the large difference between the likely mass and extent of
galaxy groups compared to isolated galaxies should enable us to
distinguish between the group-member and the isolated elliptical
hypotheses for \object{NGC1600} through a dynamical study of the other
neighbouring galaxies.

The presence of X-ray emission around \object{NGC1600} can provide an estimate
of the total mass of the gal\-axy, assuming hydrostatic equilibrium
within the gas.  Fukazawa et al. (2006), analyzing Chandra
observations of a number of elliptical galaxies, determined a
mass-to-light ratio for \object{NGC1600}, at the effective radius of 14.5 kpc,
of 10.58$M_\odot/L_\odot$, corresponding to a total mass of
approximately $10^{12}M_\odot$. This is higher than that of many
galaxies in their sample.

An investigation of the spatial distribution of galaxies around
\object{NGC1600} may also help in distinguishing between the isolated and group
hypotheses. SMG found a weak excess of galaxies around isolated
elliptical galaxies out to at least 500 kpc, with an exponential slope
of $-0.6 \pm 0.2$. This is similar to that found for poor groups and
around individual galaxies but is less steep than that found for
clusters (e.g. Hansen et al. 2005).  The number density of bright
galaxies around isolated galaxies, however, is much lower than that
found around groups and clusters.

\section{Observations and data reduction}

To obtain an estimate of the radial number density profile of galaxies
around \object{NGC1600} we have used the APM scans of UKST sky survey plates of
a circular area of radius 1.5 degrees (corresponding to a scale of
1.65 Mpc at the distance of \object{NGC1600}) surrounding this galaxy. Only
those objects brighter than 20th magnitude in $B$ and classified as
galaxies in both $B$ and $R$ were selected.  \object{NGC1600} lies within 1 degree
of the edge of plate 764 in the survey and thus the adjoining plate,
765, must be matched in both position and magnitude. Using the TOPCAT
package in the Starlink suite of astronomical data reduction packages
we have used the matching objects detected by the APM in the overlap
region of plate 764 and 765 to determine the accuracy of matching both
in magnitude and position. The average magnitude difference between
galaxies that lie in the overlap region that are detected in both
plates is less than 0.05 magnitudes and thus no magnitude difference
is assumed. In addition, the number of matched objects varies by less
than 5\% with varying the angular distance between 1 and 10 arcsec for
those objects which are classified as matched. This indicates that the
positional accuracy of the scans is less than 1 arcsec.  However, due
to several errors such as misclassification, vignetting and magnitude
errors for bright galaxies, inconsistencies will arise between the
galaxy sample derived from each plate, even though the mean density is
almost identical and the magnitudes and positions agree for duplicate
objects. The plates cannot therefore be merged as this leads to an
increase in the number density of galaxies within the overlap
region. Thus, to overcome this problem and reduce vignetting problems
at the edge of the plate, half the overlap region was taken from plate
764 and half from plate 765.

The APM scans, as in all automatic object detection methods, have
difficulty with not only star-galaxy separation at faint magnitudes
but also the correct selection of objects within the haloes of bright
stars and galaxies.  For example, diffraction spikes are quite often
split into several individual objects and their non-circular nature
often leads to such structure being classified as non-stellar. There
are several bright stars with diffraction spikes on the UKST plates of
the field around \object{NGC1600}. These are not in the immediate neighbourhood
of the central galaxy. However, to ensure that the effect of
misclassification and extraneous objects was insignificant we have
overlain the positions of the objects detected by the APM over the
UKST image for a visual inspection of the detected objects.  No
objects were found corresponding to diffraction spikes from bright
stars.  The star-galaxy separation technique used by the APM is also
not 100\% accurate and is heavily dependent upon the magnitude of the
objects, becoming highly significant beyond $B=20$ (e.g. Maddox et
al. 1990).  Thus although it is likely that some objects are
mis-classified, by limiting the sample to $B=20$ the percentage of
misclassifications is small and this is confirmed by a visual
inspection of the detected objects. A study of the detected objects
superimposed on the UKST Sky Survey plates showed that several of the
bright galaxies in the field of \object{NGC1600} were not in the catalogue
derived from the APM scans.  Many of these objects were classified as
merged objects by the APM.  To estimate the effect of these missed
galaxies on the determination of the radial number density profile,
these objects were added by eye, discounting faint objects that the
eye has difficulty classifiying and that may also fail the magnitude
cut-off. This subjective technique introduces more errors into the
analysis but will lead to some estimate of the true errors in the APM
number densities.

For the dynamical study, all galaxies within a 25 arcminute radius of
\object{NGC1600} were selected from the APM catalogue (e.g. Lewis and Irwin
1996) of the UKST survey plates.  Each object had to be detected as a
galaxy in both $B$ and $R$ to be selected and a further visual check was
made to ensure the correct identification as a galaxy. A magnitude cut
of $R=20$ was made to facilitate the possibility of getting a
redshift. As mentioned above, several bright galaxies were not in the
APM galaxy catalogue due to misclassification and these were added to
the sample by hand.

The field was observed during the nights of 7th January 2003 and 12th
January 2005 using the AUTOFIB2/WYFFOS multi-fibre spectrograph on the
4.2-m William Herschel telescope (WHT) on the island of La
Palma. During the 2003 run the large, 2.7arcsec, fibres were used to
maximise the signal from the extended sources whereas in 2005 only the
small, 1.6 arcsec, fibres were available. Additionally, in 2003 the
CCD used only enabled 120 fibres to be used whilst in 2005 the CCD had
been replaced by a two 2k x 2k EEV CCD mosaic enabling the number of
fibres available for objects to be increased to 150. The smaller CCD
used in 2003 suffered from contamination of neighbouring spectra when
bright objects were observed. As most of the galaxies observed here
were faint this was not a significant problem.

The $\tt AF2\_configure$ program was used to configure the
observations, with the brighter galaxies, and those galaxies within
the un-vignetted field of 20 arcminutes radius, preferentially
selected for observation.  With the limitations imposed by the
instrument it was impossible to observe all galaxies in the field
during one run. At least 4 stars in the field were selected for
guiding through the fiducial bundles, and unused fibres were placed on
areas of the background field for sky-subtraction.  The data from the
2003 run were reduced and the galaxies for which no redshift was
obtained, together with a selection of previously unobserved galaxies,
were selected for the 2005 run.  On both runs, a wavelength range of
approximately 3850\AA\ to 5450\AA\ was covered, with a pixel size of
0.4\AA.  Wavelength calibration was warranted by frequent observations
of a neon and helium arc.  Seeing during the 2003 run was typically
about 1 arcsec and a total of 5400 seconds integration time was
obtained on the field. In 2005 the seeing was 1.4 arcsec on average
with 7200 seconds of integration on the field.

Due to the arrangement of the fibres on the spectrograph, the observed
spectra are not uniformly placed on the CCD but are arranged in rows
of 3, with a shift of 60 pixels in the dispersion direction between
three consecutive fibres. This leads to complications in the data
reduction and so the data were reduced using the Observatory-supplied
wyffosREDUC IRAF package. This package bias-subtracts, extracts the
fibres, and subtracts the sky in an automated fashion. However, to
ensure the reduction was satisfactory, each step was visually checked.
As the sky-subtraction in this package is not ideal inspections of the
spectra were made to ensure satisfactory sky subtraction, with regions
around possible sky emission lines removed. Redshifts were obtained
using the IRAF xcsao package, with cross-correlation against a range
of templates.

\section{Results and Discussion}

Figure 1 shows the radial number density profile of galaxies centred
on \object{NGC1600} estimated from the APM scans of the UKST sky survey plates
of the region.  There is a decline in the number density with radius
and a strong suggestion that the background population has been
reached at a radius of about 1.3 Mpc. The errors displayed are
calculated assuming Poissonian statistics and are thus are likely to
be an underestimate.  Misclassification and problems with measuring
bright galaxies will almost certainly increase the error estimates,
whilst in the inner 150 kpc the small number of detected galaxies and
the blanking factor due to the presence of \object{NGC1600} will lead to
greater uncertainties. The difference between the profile determined
purely by the APM detections and that with galaxies missed by the APM
but selected by eye included lies within the errors.

\begin{figure}
    \includegraphics[width=9.cm]{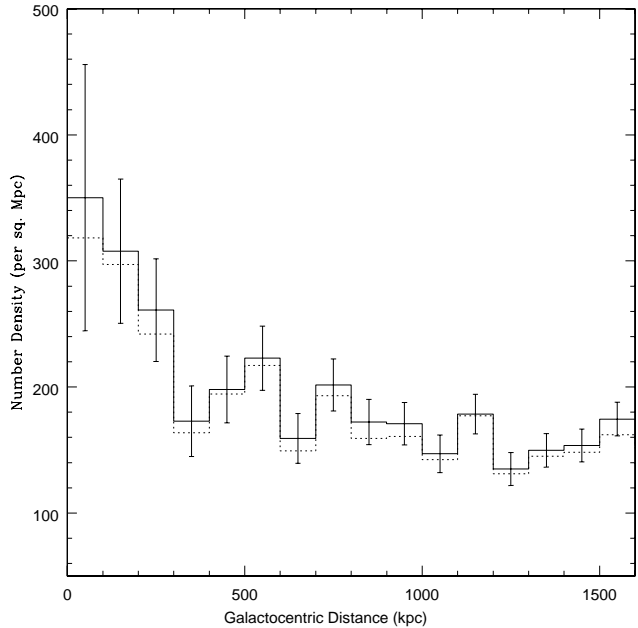}
\caption{Projected density of galaxies brighter than $B=20$ around NGC1600.
The solid line includes galaxies selected by eye, whilst the dotted line
is only those galaxies detected by the APM. At the redshift of
NGC1600, $B=20$ is equivalent to $M_B=-14$.}
\end{figure}

Figure 1 clearly shows a decrease in the number density of galaxies
with galactocentric distance.  The background galaxy population
density appears to be reached at a galactocentric distance of about
1.3 Mpc, for a number density of approximately 150 galaxies per square
Mpc. Fitting an exponential to this distribution gives a slope of
$-0.9\pm 0.3$. Although slightly steeper than the $-0.6\pm0.2$ slope
found by SMG for their total sample of isolated ellipticals, they
agree within the errors. However, the slope also agrees within the
errors with the slope of $-1.1$ found for groups and clusters by Hansen
et al. (2005) and also for an isothermal distribution.  From their
study of a sample of 10 isolated ellipticals, Reda et al. (2004) claim
that only the faint dwarf galaxies, with $M_R \gtrsim -15.5$, show any
clustering around the central galaxy. At the distance of \object{NGC1600}, the
limiting magnitude of our sample corresponds to an absolute $B$
magnitude of $-14$ and therefore we would expect to see the excess of
neighbours. By increasing the limiting magnitude to $R=18$,
corresponding to an absolute magnitude of $M_R = -16$ the number
densities decrease significantly and the error bars therefore
increase. Figure 2 shows the resultant density profile for this
brighter magnitude limit. Although there is evidence of a small excess
of galaxies at small galactocentric distances, and these objects are
seen by visual inspection of the field, the error bars are large
enough such that a uniform distribution, as suggested by Reda et
al. (2004), is not ruled out.

\begin{figure}
    \includegraphics[width=9.cm]{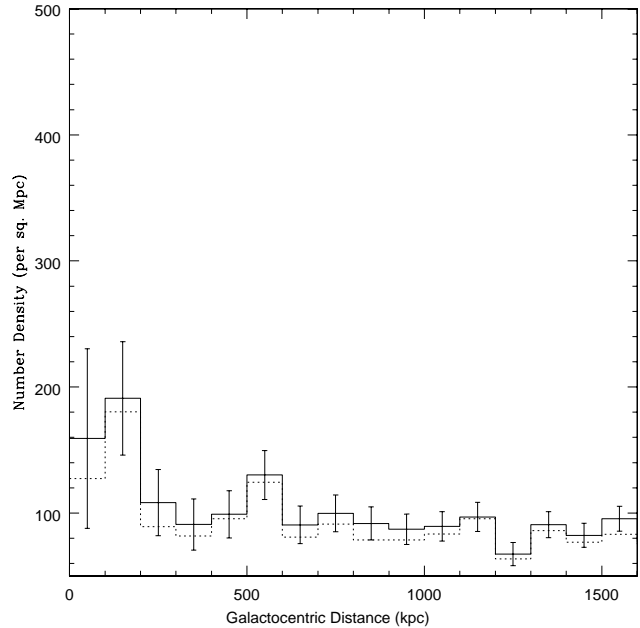}
\caption{Projected density of galaxies brighter than $R=18$ around NGC1600.The
solid line includes those galaxies selected by eye whilst the dotted line
represents only those galaxies detected by the APM.}
\end{figure}

The radial density profile thus indicates that there is a significant
population of faint ($M_B \gtrsim -15$) galaxies in the neighbourhood of
\object{NGC1600} similar to that found, in general, from a sample of isolated
ellipticals by SMG. These results are in agreement with the results of
Reda et al. (2004) who suggest that apparently isolated galaxies do
not have a large number of bright galaxies in their neighbourhood but
could be surrounded by fainter galaxies.  The total excess of
galaxies\footnote{Compared to the background reached at approximately
1.3 Mpc, with a number density of approximately 150 $B<20$ galaxies
per square Mpc, as seen in previous paragraphs.} within 500 kpc down
to $M_B = -14.6$ is approximately 55, in good agreement with the
average for the sample of 10 galaxies of SMG. From the radial number
density profiles, therefore, \object{NGC1600} does not appear unusual with
respect to other isolated early-type galaxies, with an extended
population of neighbouring faint galaxies but a lack of bright
companions.

Galaxies in the field of \object{NGC1600} with measured redshifts are shown in
Table 1. Previously measured redshifts are also given, together with
those galaxies listed in NED and HyperLeda with redshifts between 3700
and 5700 $\rm{km\ s}^{-1}$ and within a projected distance of 57
arcminutes of \object{NGC1600} (corresponding to a physical projected distance
of 1 Mpc). Note that this latter list includes members of the galaxy
group Zw0430.8-0424B catalogued by Zwicky et al. (1961-1968).

\begin{table*}
\begin{center}
\caption{Galaxies in the NGC1600 area observed in this work (upper
part) or in the literature.}
\begin{tabular}{ccccrl} \tableline \tableline

&&&&&\\

RA (J2000)&Dec (J2000)&B mag&$v$&Error&Comments\\
\tableline
&&&&&\\

$04:31:39.86$&$-05:05:09.6$&11.93&4715&16&NGC1600, parent\\
&&&&&$\rm 4681\pm 8$\tablenotemark{a}, \,$\rm 4739\pm 33$\tablenotemark{b}  \\
$04:30:04.25$&$-05:00:35.6$&15.32&4196&10&\\
$04:30:09.00$&$-05:21:16.9$&16.39&5321&9&\\
$04:30:23.18$&$-04:53:33.0$&16.40&3729&34&\\
$04:30:42.73$&$-04:52:13.4$&14.78&4009&10& IC373\\
$04:30:58.66$&$-05:13:56.4$&19.77
&4704&27&\\
$04:31:07.69$&$-05:01:09.3$&18.25&5209&30&\\
$04:31:15.82$&$-04:57:03.2$&16.16&4947&9&\\
$04:31:28.66$&$-05:16:28.9$&16.36&4422&10&\\
$04:31:41.70$&$-05:03:37.4$&14.80&4887&12&NGC1601, $\rm 4997\pm 56$\tablenotemark{c}\\
$04:31:43.18$&$-04:59:14.3$&16.83&5704&19&\\ 
$04:31:47.10$&$-05:16:07.2$&16.74&4399&21&\\
$04:31:49.94$&$-05:05:39.8$&15.15&4990&13&NGC1603, $\rm 4972\pm 14$\tablenotemark{d}\\
$04:31:58.55$&$-05:22:11.6$&14.46&4473&13&NGC1604, $\rm 4544\pm 25$\tablenotemark{e}\\
$04:32:03.30$&$-05:01:57.0$&15.91&4802&23& NGC1606 \\
$04:32:10.39$&$-05:08:12.1$&16.14&4739&12&\\
$04:32:24.86$&$-05:15:50.1$&16.93&4562&26&\\
$04:32:25.40$&$-05:11:18.9$&16.13&5092&25&\\
$04:32:30.37$&$-05:09:18.7$&16.13&4630&7&\\
\tableline
\tableline
&&&&&\\
$04:28:18.8$&$-05:10:44$&14.35&4267&25&NGC1580\\
$04:28:25.7$&$-04:37:50$&$>15$&4680&60&\\
$04:28:26.0$&$-04:33:49$&15.75&4846&39&\\
$04:30:34.7$&$-05:31:54$&17.72&4908&60&\\
$04:30:51.6$&$-05:47:54$&13.72&4329&4&NGC1594\\
$04:31:12.5$&$-05:31:44$&16.26&5202&60&\\
$04:31:38.8$&$-04:35:18$&14.36&4008&8&NGC1599\\
$04:31:52.1$&$-05:45:25$&15.47&4121&41&\\
$04:32:03.1$&$-04:27:38$&14.51&4255&3&NGC1607\\
$04:32:08.4$&$-04:12:43$&15.74&4564&60&\\
$04:33:01.0$&$-04:11:19$&16.82&4904&15&\\
$04:33:05.9$&$-04:17:51$&14.40&4261&15& NGC1611\\

\tableline

\end{tabular}
\tablenotetext{a}{NED, from Collobert et al. 2006.}
\tablenotetext{b}{HyperLeda average.}
\tablenotetext{c}{NED, from de Vaucouleurs et al. 1991.}
\tablenotetext{d}{NED, from Simien and Prugniel 2000.}
\tablenotetext{e}{NED, from Huchra et al. 1993.}
\tablecomments{At the lower part of the table, we list 
other galaxies within a projected separation of 1 Mpc
and 1000 $\rm{km\ s}^{-1}$ of NGC1600 but not observed by us. 
Data from HyperLeda.}

\end{center}
\end{table*}

With a total of 30 known galaxies within 1000 $\rm{km\ s}^{-1}$ of
\object{NGC1600}, our sample indicates that a very significant population of
galaxies exists in the physical neighbourhood of \object{NGC1600}, as expected
from previous observations by other authors and also the photometric
study described above. The large number of measured redshifts allows
an estimate of the velocity distribution of the galaxies in the field
of \object{NGC1600} to be determined, together with an estimate of the mass.
Figure 3 shows the relative velocity distribution of the 30 companions
with respect to \object{NGC1600}. Also shown is the best-fit Gaussian,
with a mean relative velocity of $-85$ $\rm{km\ s}^{-1}$ and a velocity
dispersion of 435 $\rm{km\ s}^{-1}$.

\begin{figure}
    \includegraphics[width=9.cm]{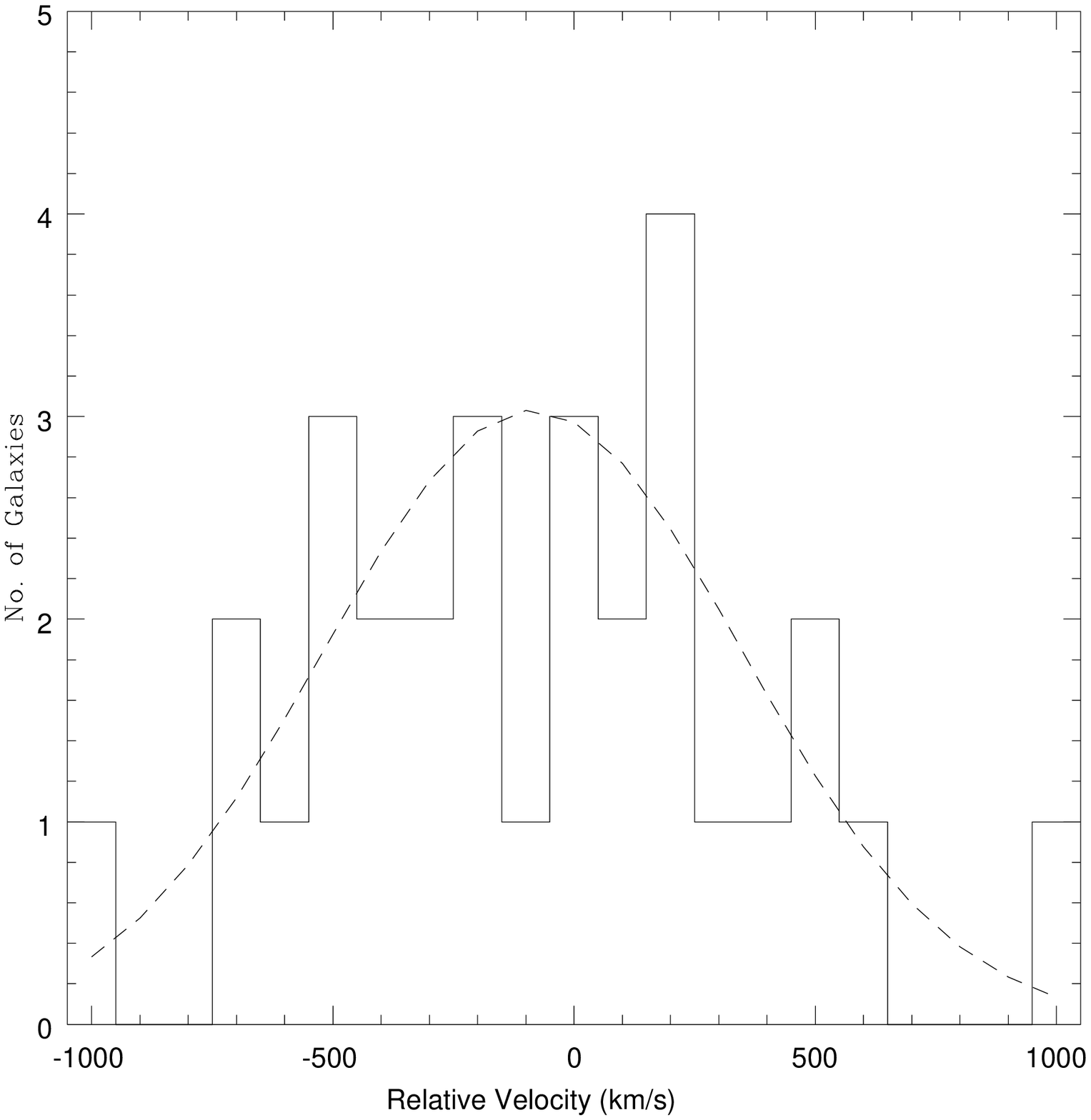}
\caption{Histogram showing the velocity distribution of the galaxies
in the NGC1600 group, relative to NGC1600 itself. The dashed line
corresponds to the best-fit Gaussian, with $v_{\rm{cent}}=-85 \, \rm{km\
s}^{-1}$ and $\sigma_v=435 \, \rm{km\ s}^{-1}$.}
\end{figure}

In order to obtain a more precise and less model-dependent measurement
of these parameters we have also used the ROSTAT code to estimate the
mean redshift of the group and its radial velocity dispersion (Beers,
Flynn \& Gebhardt 1990), with the usual cosmological correction and
the correction for velocity errors given by Danese, de Zotti \& Tullio
(1980). Given that there is a large number of redshifts available, the
biweight estimators were used for both the location and scale (Beers
et al. 1990). Errors were obtained in all cases by jackniving of the
biweight. Using this machinery we measure the group's central location
at $4634 \pm 79 \, \rm{km\ s}^{-1}$, with a radial velocity dispersion
value of $429 \pm 57 \, \rm{km\ s}^{-1}$.

The redshift of \object{NGC1600} lies $1\sigma$ away of the system central
velocity, and thus it appears that \object{NGC1600} is not at the dynamical
centre of the group, even though it is the brightest member by over 2
magnitudes. This is in agreement with Sivakoff et al. (2004) who found
that the X-ray emission is centred slightly to the north-east of
\object{NGC1600}, also suggesting that the galaxy is not at the centre of the
gravitational potential. An estimate of the geometry of the system can
be ascertained from Figure 4, where we show the positions and
velocities relative to \object{NGC1600} of the galaxies for which a redshift
has been measured. 

\begin{figure*}
\begin{center}
    \includegraphics[width=13.cm]{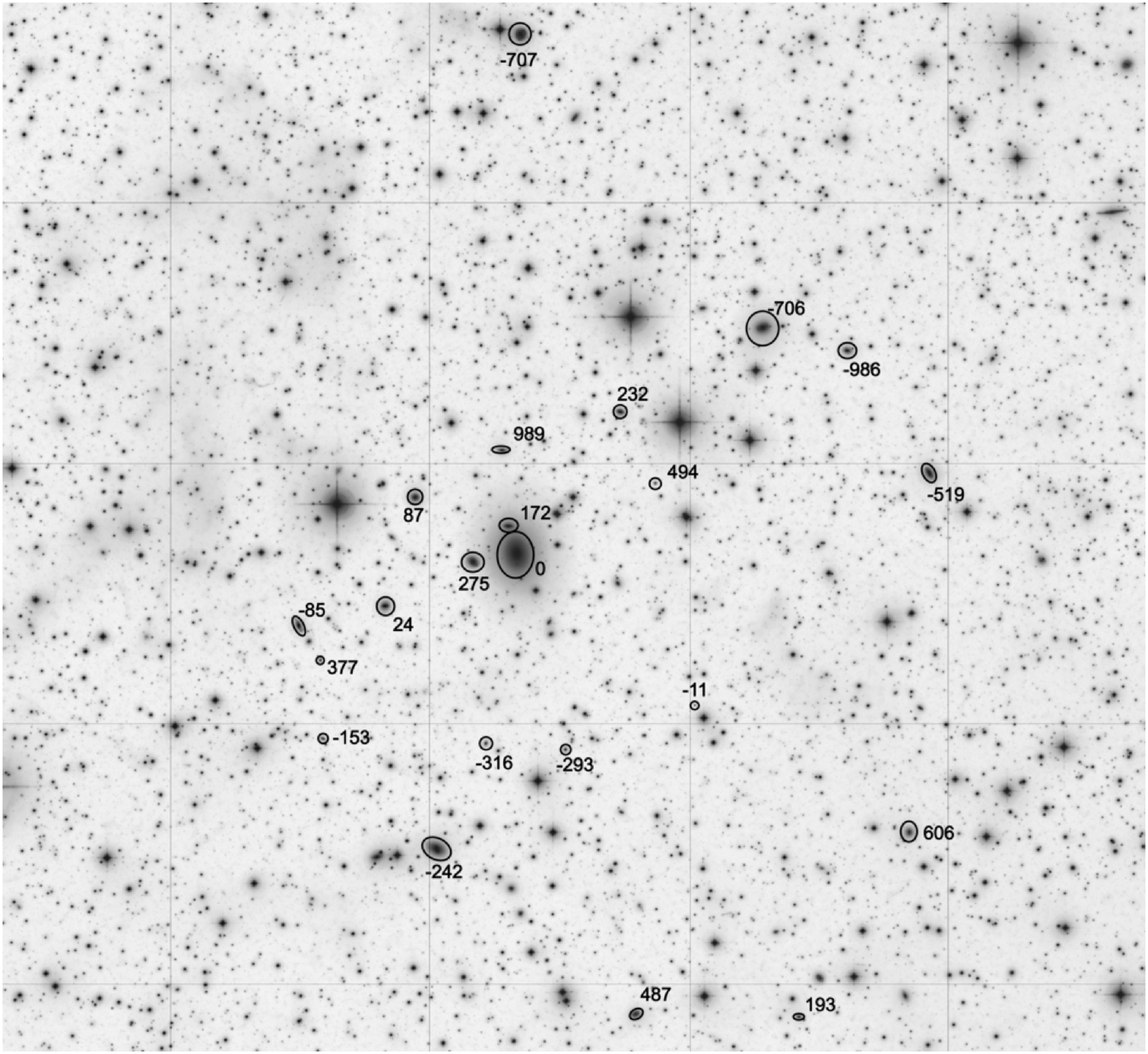}
\caption{UKST Sky Survey 1 degree square image of the NGC1600
  field. Those galaxies for which a redshift has been measured are
  marked, and their velocity relative to NGC1600 has been labeled. Notice 
that some galaxies in Table 1 are outside this field.
Use of these images is courtesy of the UK Schmidt Telescope (copyright in
which is owned by the Particle Physics and Astronomy Research Council of
the UK and the Anglo-Australian Telescope Board) and the Digitized Sky Survey
created by the Space Telescope Science Institute, operated by AURA, Inc, for
NASA, and is reproduced here with permission from the Royal Observatory
Edinburgh.}
  \end{center}
\end{figure*}

On the other hand, the velocity dispersion of 429 $\rm{km\ s}^{-1}$
would imply a bolometric X-ray luminosity $L_X = 2.7 \times 10^{43}
{\rm erg\ s}^{-1}$, as in the empirical relation of Ortiz-Gil et
al. (2004), which was measured from a sample of 171 clusters drawn
from the REFLEX catalogue. Sivakoff et al. (2004) measure an X-ray
flux which is two orders of magnitude smaller\footnote{REFLEX
luminosities are bolometric, while Sivakoff et al. (2004) measure only
in the soft band. However, the corrections from soft to bolometric
represent a factor between 1.1 and 2 for this kind of groups/clusters
(Bohringer et al, 2004, Table 5).}. Although the $L_X-\sigma$ relation
is not well defined, \object{NGC1600} is more than $3\sigma$ away from 
the expected value, and hence lies significantly outside the scatter.

A velocity dispersion value of 429 $\rm{km\ s}^{-1}$ for an elliptical
galaxy is very high, placing it well off the Faber-Jackson
relationship (Faber and Jackson 1976). Hau and Forbes (2006) measured
the radial distribution of the velocity dispersion for the Reda et
al. (2004) sample of isolated galaxies. They find no galaxy with a
similar velocity dispersion to that found here, with an increase in
the $V/\sigma$ with radius and a general decrease of velocity
dispersion with radius, out to the effective radius. Velocity
dispersions of loose groups are found to be of the order of a few
hundreds of kilometers per second (e.g. Ramella et al. 1995), with a
value of 429 $\rm{km\ s}^{-1}$ being at the upper limit for loose
groups, but at the lower range for clusters of galaxies.  Making the
somewhat arbitrary assumption that the group can be approximated by an
isothermal sphere it is possible to derive an upper limit for the mass
of the group from $M(R) = (\pi {\sigma}^2 R)/G$ (Binney and
Merrifield, 1998). A simple application of this formula to the
\object{NGC1600} `group' gives a mass of approximately $2 \times
10^{14} M_\odot$, typical of the richer groups or poorer
clusters. Although only 30 satellites have been detected, we can also
use the estimate of Bahcall and Tremaine (1981) to derive an estimate
of the mass of the central object. Assuming that the galaxies are
distributed uniformly on radial and circular orbits, the mass of the
central object is given by $M = 48\langle r\Delta v^2/2G
\rangle/\pi$. Applying this formula to the group data gives a mass
estimate of $1.5 \times 10^{14} M_\odot$, in agreement with the
isothermal sphere determination.  Zaritsky and White (1994) have shown
from extensive modelling that errors in such an estimate are not large
as long as interlopers are excluded. However, the presence of
interlopers can have a serious effect on mass determinations of
galaxies and clusters (e.g. Chen et al. 2006). To estimate the effect
of interlopers on our sample, we can remove from the sample those
galaxies whose relative velocities are greater than 900 (500)$\rm km
s^{-1}$. This would drop the value of the mass from $1.5 \times
10^{14} M_\odot $ to $1.2 \times 10^{14} (7.9 \times 10^{13}) M_\odot
$. Further elimination of those galaxies at projected distances
greater than 1Mpc from NGC 1600 would drop the mass estimate even
further to $2.1 \times 10^{13} M_\odot$, but for a sample of only 13
galaxies. 
This lower estimate is typical of loose groups (e.g.  Parker
et al. 2005) and is perhaps larger than that expected from dynamical
observations of the inner regions of elliptical galaxies such as found
from the study of planetary nebulae (e.g. Romanowsky et
al. 2003). Fuzakawa et al.  (2006) have also used the X-ray emission
to derive an estimate of 10.58 for the mass-to-light ratio for the NGC
1600 `group' out to the effective radius of 13.8 kpc, giving a mass of
approximately $10^{12} M_\odot$ at this distance. Extrapolating this
estimate to larger radii is subject to considerable error but, if the
radial mass-to-light ratio profile for NGC 1600 is similar to that
found for other ellipticals by Fuzakawa et al. (2006), the mass
determination from the X-ray emission is not in major disagreement
with that from our dynamical study. However, the variations 
associated with the different methods suggest that our estimates 
of the group  mass are still subject to an uncertainty of 
about a factor of two.

The results of the number density and radial velocity studies allow us
to probe further our definition of an isolated galaxy. The former
study indicates that there is a population of much fainter galaxies
surrounding \object{NGC1600}, with a small number of brighter galaxies in the
immediate vicinity which are not seen in the sample of Reda et
al. (2004). However, these bright galaxies have a negligible effect on
the overall number density profile around \object{NGC1600}. The profile is
therefore not significantly different from that found from other
isolated early-type galaxies. The published photometric properties are
also similar to those for isolated ellipticals, with little evidence
of recent merging. The dynamical study however, together with the
X-ray data, indicates that \object{NGC1600} is not sitting at the centre of
the potential and is surrounded by a massive halo extending out to
several hundred kiloparsecs. This would normally be taken as evidence
that \object{NGC1600} is a member of a group of galaxies.

In any case, a simple comparison of the $(J-K)$ {\it vs} $K$ colour diagram of
the \object{NGC1600} group of galaxies with that of the Coma Cluster (Figure 5)
shows that there are similarities between both fields. The main differences
are, of course, the large difference in richness and the already mentioned
magnitude gap between \object{NGC1600} and the second $K$-band brightest galaxy in
the group (NGC1611), which is much larger than in Coma.

\begin{figure}
    \includegraphics[height=9.cm,angle=-90]{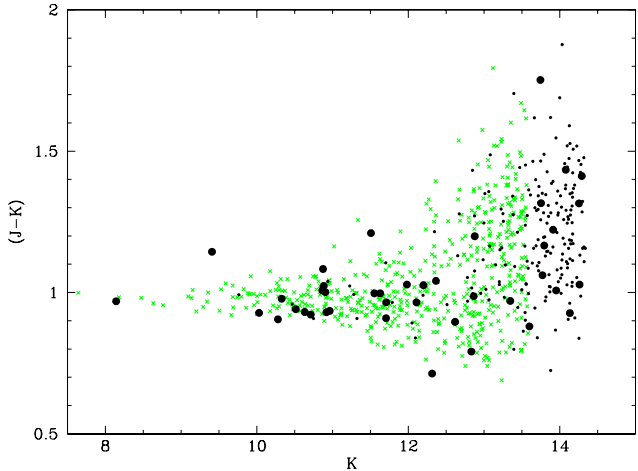}
\caption{$JK$ colour-diagram of the NGC1600 field. The dark large circles
  correspond to objects with redshifts that link them to NGC1600, whereas
  the small circles are field objects without a redshift. As a comparison,
  the crosses mark the positions of galaxies in the Coma cluster field,
  corrected to the distance of NGC1600.}
\end{figure}

There are several similarities between \object{NGC1600} and the fossil groups of
Jones et al. (2003): the large magnitude difference between it and its
neighbours, the presence of a surrounding faint population, and the
presence of extended X-ray emission. In addition, the absolute magnitude of
\object{NGC1600} ($M_R \sim -23.0$) lies within the range of luminosities found for
other fossil group central galaxies, although at the fainter end of the
range. Moreover, the measured velocity dispersion and absolute $B$-band
magnitude position \object{NGC1600} perfectly in the $\sigma_V$ {\it vs} $M_B$ relation
for fossil groups, as described by Khosroshahi et al (2006). If the scaling
properties of the fossil groups as measured by D'Onghia et al (2005) in
their simulations apply, \object{NGC1600} would have been already in place (that is,
more than 50\% of the mass would already have been assembled) 7-8 Gyrs ago,
which matches the available age estimates. However, the X-ray properties of
the \object{NGC1600} group are very different from those expected, with a much
smaller extent and a much lower luminosity (Sivakoff et al 2004). Thus,
\object{NGC1600} could be taken as a new class of fossil group, where the central
galaxy has grown from the merger of several fainter group members but is
lacking the relatively intense X-ray emission normally associated with
loose groups. Thus it could be the result of a merger of a poor group,
where the X-ray emission is expected to be much weaker.  The presence of
dust and emission regions within the galaxy support this merger hypothesis.
NGC 1132, an isolated elliptical galaxy studied by Mulchaey and Zabludoff (1999),
has similar properties to \object{NGC 1600}, with a faint extended X-ray envelope and  
surrounding excess population of dwarf galaxies, although the dynamical properties
of the group are uncertain. Thus \object{NGC1600} 
may not be the only example of such a past merger.

\section{Conclusion}
Although \object{NGC1600} is clearly surrounded by a number of other
galaxies, the large magnitude difference suggests that it can be
treated as an isolated, field, elliptical. Analysis of the APM scans
of UKST Sky Survey plates shows that there is a clear excess of
galaxies within the neighbourhood of \object{NGC1600}. In partial agreement
with Reda et al. (2004) the excess is most clearly evident for the
fainter galaxies, with galaxies brighter than $M_R = -16$ not
showing such a clear excess. The number of galaxies, together with
their number density profile, around \object{NGC1600} is in agreement with
the results of SMG indicating that this galaxy is not unique amongst
isolated early-type galaxies.

Velocity measurements of galaxies within the field show a large
excess at similar redshifts to \object{NGC1600}. The dynamics of these
galaxies indicates that the mass of the group is at least $10^{13}
M_\odot$, a value typical of poor groups. Combining the data from
these and previous studies indicates that \object{NGC1600} has many
similarities to the fossil groups first detected in X-rays, although
there are several differences, most notably in the X-ray properties.
We therefore suggest that \object{NGC1600} is a new type of fossil group, the
result of the merger of a loose group of galaxies, with the last
merger occuring over 4 Gyr ago.

The similarity between \object{NGC1600} and other galaxies in the SMG sample
of isolated ellipticals suggests that many such objects may be the
result of the merger of loose groups. This will only be determined by
a detailed photometric and dynamical study of a larger sample of
isolated galaxies.  As the majority of galaxies lie in loose groups,
including our own, such an investigation would lead to a greater
understanding of galaxy evolution.

\acknowledgements

The authors acknowledge the comments of our anoymous referee, which
helped in clarifying the content of this paper.  We also thank Elena
D'Onghia for her helpful comments. VJM, AFS, FJB, and AOG acknowledge
support from the Spanish Minsiterio de Educaci\'on y Ciencia (MEC) through 
project AYA2006-14056 (including FEDER). AFS
acknowledges support from the Spanish MEC through a ``Ram\'on y
Cajal'' contract. We thank Vicent Peris for his help in processing the
DSS image of \object{NGC1600} for Fig. 4 using PixInsight.  This work is based
on observations obtained at the 4.2-m William Herschel Telescope
(WHT), operated on the island of La Palma by the Isaac Newton Group in
the Observatorio del Roque de Los Muchachos of the Instituto de
Astrof\'{\i}sica de Canarias.  The authors acknowledge the assitance
given by the support astronomers at WHT.  This publication makes use
of data products from the Two Micron All Sky Survey, which is a joint
project of the University of Massachusetts and the Infrared Processing
and Analysis Center/California Institute of Technology, funded by the
National Aeronautics and Space Administration and the National Science
Foundation. We acknowledge the usage of the HyperLeda database 
(\url http://leda.univ-lyon1.fr). 

{\it Facilities:} \facility{ING:Herschel ()}


\begin{thebibliography}{}
\bibitem[a]{a}Bahcall, J.N. and Tremaine, S., 1981, ApJ, 244, 805
\bibitem[b]{b}Baiesi-Pillastrini, G.C., Palumbo, G.C.C., Vettolani, G., 1984, A\&ASS, 56 ,363
\bibitem[ba]{ba}Beers T.C., Flynn K., Gebhardt K., 1990, AJ, 100, 32
\bibitem[c]{c}Bender, R., Saglia, R.P. and Gerhard, O.E., 1994, MNRAS, 269, 785
\bibitem[d]{d}Binney, J., and Merryfield, M., 1998, Galactic Astronomy, (Princeton U Press)
\bibitem[e]{e}Birkinshaw, M. and Davies, R. L. 1985, ApJ, 291, 32
\bibitem[ea]{ea}Bohringer et al., 2004, A\&A, 425, 367
\bibitem[f]{f}Chen,J., Kravtsov, A.V., Prada, F., Sheldon, E.S., Klypin, A.A., Blanton, M.R.,
Brinkmann, J. and Thakar, A.R., 2006, ApJ, 647, 86
\bibitem[g]{g}Colbert, J.W., Mulchaey, J.S. and Zabludoff, A.I., 2001. AJ, 121,808
\bibitem[g1]{g1} Collobert, M., Sarzi, M., Davies, R.~L., Kuntschner, 
H., \& Colless, M.\ 2006, MNRAS, 370, 1213 
\bibitem[ga]{ga}Cypriano, E.~S., Mendes de Oliveira, C.~L., 
\& Sodr{\'e}, L.~J.\ 2006, AJ, 132, 514 
\bibitem[gb]{gb}Danese L., de Zotti G., di Tullio G., 1980, A\&A, 82, 322
\bibitem[gb1]{gb1} de Vaucouleurs, G., de Vaucouleurs, A., Corwin, H.~G., 
Jr., Buta, R.~J., Paturel, G., \& 
Fouque, P.\ 1991, Volume 1-3, XII, Springer-Verlag, Berlin.  
\bibitem[h]{h}Denicolo, G, Terlevich, R., Terlevich, E., Forbes, D.A., Terlevich, A. and
Carrasco, L., 2005, MNRAS, 356, 1440
\bibitem[i]{i}D'Onghia, E., Sommer-Larsen, J., Romeo, A.D., Burkert, A., Pedersen, K.,
Portinaril, L., and Rasmussen, J., 2005, ApJ, 630, L109
\bibitem[j]{j}Faber, S.M. and Jackson, R.E., 1976, ApJ, 204, 668
\bibitem[k]{k}Faber, S. M., Wegner, G., Burstein, D., Davies, R. L., Dressler, A.,
Lynden-Bell, D. and Terlevich, R. J. 1989, ApJS, 69, 763
\bibitem[l]{l}Ferrari, F., Pastoriza, M. G., Macchetto, F. and Caon, N. 1999, A\&AS, 136, 269
\bibitem[m]{m}Fukazawa, Y., Botoya-Nonesa, J.G., Pu, J., Ohto, A. and Kawano, N., 2006, ApJ,636,698
\bibitem[o]{o}Hansen, S.M., McKay, T.A., Wechsler, R.H., Annis, J., Sheldon, E.S. and
Kimball, A., 2005, ApJ, 633, 122 
\bibitem[p]{p}Hau, G.T.K. and Forbes, D.A., 2006, MNRAS, 371, 633
\bibitem[p1]{p1} Huchra, J., Latham, D.~W., da Costa, L.~N., Pellegrini, P.~S., \& Willmer, C.~N.~A.\ 1993, \aj, 
105, 1637 
\bibitem[r]{r}Lewis, G. and Irwin, M.J., 1996, Spectrum, Newsletter of the Royal 
Observatories, 12, 22
\bibitem[s]{s}Jones, L. R., Ponman, T. J., Horton, A., Babul, A., Ebeling, H. and Burke, D. J.,2003,
MNRAS,343,627
\bibitem[t]{t}Karachentsev, I.D., 2005,AJ, 129,178
\bibitem[u]{u}Khosroshahi, H.G., Ponman, T.J., and Jones, L.R., 2006, MNRAS, 372, L68
\bibitem[v]{v}Matthias, M. and Gerhard, O., 1999, MNRAS, 310, 879
\bibitem[va]{va} Mendes de Oliveira, C.~L., Cypriano, E.~S., 
\& Sodr{\'e}, L.~J.\ 2006, AJ, 131, 158 
\bibitem[w]{w}Mulchaey, J.S. and Zabludoff, A.I., 1999,ApJ, 514,133
\bibitem[wa]{wa}Ortiz-Gil, A., Guzzo, L., Schuecker, P., Bohringer, H., Collins, C.A.
2004, MNRAS, 348, 325O
\bibitem[x]{x}Parker, L.C., Hudson, M.J., Carlberg, R.G. and Hoekstra, H., 2005, ApJ, 
634, 806
\bibitem[y]{y}Reda, F.M., Forbes, D.A., Beasley, M.A., O'Sullivan, E.J. and Goudfrooij, P.,
2004, MNRAS, 354, 851
\bibitem[z]{z}Ramella, M., Geller, M.J., Huchra, J.P. and Thorstenson, J.R., 1995, AJ,
109, 1458
\bibitem[aa]{aa}Romanowsky, A.J., Douglas, N.G., Arnaboldi, M., Kuijken, K.,
Merrifield, M.R., Napolitano, N.R., Capaccioli, M. and Freeman, K.C., 2003,
Science, 301, 1698
\bibitem[aaa]{aaa} Santos, W.~A., Mendes de Oliveira, C., \&
  Sodr{\'e}, L.~J.\ 2007, AJ, 134, 1551 
\bibitem[ab0]{ab0} Simien, F., \& Prugniel, P.\ 2000, A\&AS, 145, 263 
\bibitem[ab]{ab}Sivakoff, G.R., Sarazin, C.L. and Carlin, J.L., 2004, ApJ, 617, 262
\bibitem[ac]{ac}Smith, R.M., Mart\'{\i}nez, V.J. and Graham, M.J., 2004, ApJ, 617,1017 (SMG)
\bibitem[ad]{ad}Terlevich, A. I. and Forbes, D. A. 2002, MNRAS, 330, 547
\bibitem[ae]{ae}Trager, S. C., Faber, S. M., Worthey, G. and Gonzalez, J. J. 2000, AJ, 119, 1645
\bibitem[af]{af}Trinchieri, G. and di Serego Alighieri, S. 1991, AJ, 101, 1647
\bibitem[ag]{ag}Zaritsky, D., Smith, R.M., Frenk, C.S. and White, S.D.M., 1993, ApJ,
405, 464
\bibitem[ah]{ah}Zaritsky, D., Smith, R.M., Frenk,C.S. and White, S.D.M., 1997, ApJ,
478, 39
\bibitem[ai]{ai}Zaritsky, D. and White, S.D.M., 1994, ApJ, 435, 599
\bibitem[aj]{aj}Zwicky, F., Herzog W., Wild, P., Karopowicz, M, and Kowal D., 1961-1968,
Catalog of Galaxies and of Clusters of Galaxies (Pasadena: Caltech)
\end{thebibliography}
\end{document}